# Giant modulation of the magnetic domain size induced by an electric field

F. Ando,[1,†] H. Kakizakai,[1,†] T. Koyama,[2] K. Yamada,[1] M. Kawaguchi,[1] S. Kim,[1] K-J. Kim,[1] T. Moriyama,[1] D. Chiba,[2,*] and T. Ono[1,*]

[1]*Institute for Chemical Research, Kyoto University, Gokasho, Uji, Kyoto, 611-0011, Japan.*

[2]*Department of Applied Physics, The University of Tokyo, Hongo 7-3-1, Bunkyo, Tokyo, 133-8656, Japan.*

**The electric field (EF) effect on the magnetic domain structure of a Pt/Co system was studied, where an EF was applied to the top surface of the Co layer. The width of the maze domain was significantly modified by the application of the EF at a temperature slightly below the Curie temperature. After a detailed analysis, a change in the exchange stiffness induced by the EF application was suggested to dominate the modulation of the domain width observed in the experiment. The accumulation of electrons at the surface of the Co layer resulted in an increase of the exchange stiffness and the Curie temperature. The result was consistent with the recent theoretical prediction.**

[†]These authors contributed equally to this work
*Correspondence to: dchiba@ap.t.u-tokyo.ac.jp, ono@scl.kyoto-u.ac.jp

The electric field (EF) control of magnetism[1–21] was intensively investigated because of its potential importance for the reduction of power consumption in magnetic storage devices.[22] For the realization of EF-assisted or -induced magnetization switching,[2,8,12–14,20] the modulation of magnetic anisotropy (MA) is of great importance.[6,7,8,19] The mechanism of the EF effect on MA in 3*d*-transition metals was considered using the electron occupancy change at *d*-orbitals caused by a Fermi level $E_F$ shift and/or a change in the electronic structure near $E_F$.[7,23,24] Not only the MA modulation but also a change in Curie temperature $T_C$ was reported in a metallic Pt/Co system[10,11] as well as in ferromagnetic semiconductors.[1,3] Using ab initio calculation, the $T_C$ change due to EF application to a Pt/Co system was suggested to be explained by the modulation of the Heisenberg exchange parameter.[25] Observation of the domain wall motion[4,15–18] or the domain structure modified by an EF[26–28] is expected to be one of methods to investigate the EF effect on the magnetic parameter related to the exchange, *i.e.*, the exchange stiffness. In this paper, we report the EF effect on the magnetic domain structure in a Pt/Co system with a perpendicular magnetic anisotropy (PMA). The width of the maze domain was significantly changed by the application of an EF at a temperature slightly below $T_C$. After a detailed analysis, a change in the exchange stiffness induced by the EF application was suggested to dominate the modulation of the domain width observed in this experiment.

To observe the EF effect on the magnetic domain structure using a magneto-optical Kerr effect (MOKE) microscope with a polar-Kerr configuration, a device with a transparent gate electrode made of an InSnO (ITO)[18,26,28] was fabricated as shown in Fig. 1. The Pt/Co sample was deposited on the intrinsic Si (001) substrate with thermally oxidized layer ($SiO_2$) on top using RF sputtering. The layer structure was as follows: Ta (3.3 nm)/Pt (2.4 nm)/Co (0.27 nm)/MgO (2.0 nm) from the substrate side. The temperature $T$ dependence of the perpendicular magnetization $M_\perp$ curve of the as-deposited sample at external perpendicular field $\mu_0 H_\perp = 2.5$ mT is shown in Fig. 2(a). $M_\perp$ decreased to zero rapidly at a temperature of ~350 K, indicating that $T_C$ exists near this temperature. To observe the magnetization curves using the anomalous Hall effect as well as MOKE images, a 1.5-mm-wide wire structure with Hall probes was fabricated using photolithography and Ar-ion milling. Then the sample was covered by a $HfO_2$ (50 nm) gate insulator in an atomic layer deposition chamber. Finally, an ITO gate electrode was deposited using RF sputtering. Gate voltage $V_G$ was applied between the gate electrode and the Co layer. Here, positive $V_G$ was defined as the direction of electron accumulation at the top surface of the Co layer.

The Hall resistance $R_{Hall}$ curves for $V_G = 0$ V near the Curie temperature are shown in the inset of Fig. 2(a). The right vertical axis shows $1/\chi_{Hall}$ plot for $V_G = 0$ V, where the slope of the $R_{Hall}$ curves in the inset was used as magnetic susceptibility $\chi_{Hall}$ for each temperature. To determine the slope, the data in the range of $\mu_0|H_\perp| < 10$ mT were used. The linear fit to $1/\chi_{Hall}$ data at a higher temperature showed that paramagnetic Curie temperature $\Theta_f$ of the sample was 347 K. $\Theta_f$ determined by $1/\chi_{Hall}$ plot for $\gamma \sim 1$ was consistent with $T_C$ determined by the Arrott plot for the previous similar samples, where $\gamma$ is the critical exponent in the following relationship: $1/\chi \sim 1/(T - \Theta_f)^\gamma$ (see supplementary information in ref. 10). Thus, a temperature of 320 K, at which the main experiments were performed in this study, is the temperature slightly below $T_C$ of the present sample.

Figure 2(b) shows the $R_{Hall}$ curves for $V_G = +10$, 0, and $-10$ V, which were obtained by sweeping $\mu_0 H_\perp$ at 320 K. Square hysteresis curves originating from the PMA were observed for $V_G = +10$ and 0 V, whereas the curve for $V_G = -10$ V had a moderate shape. In addition, $R_{Hall}$, which is proportional to the saturation magnetization, and the coercivity decreased as $V_G$ was decreased. These results suggest that $T_C$ of the sample was reduced by the application of $V_G$ in the negative direction.[10,11] The $R_{Hall}$ curves for $V_G = +10$, 0, and $-5$ V at 340, 335, and 330 K, respectively are shown in Fig. 2(c). The curves completely overlapped each other, indicating that the difference in $T_C$ ($\Delta T_C$) of $V_G = +10$ V from $T_C$ at 0 V was $\sim+5$ K, while $\Delta T_C$ of $V_G = -5$ V was $\sim-5$ K.

Figures 3(a)–3(c) show the MOKE images taken under three different $V_G$s at 320 K. First, to demagnetize the sample, the sample was heated up to 330 K at $\mu_0 H_\perp = 0$ T. Then $V_G$ was applied and the temperature was reduced to 320 K. Before taking the MOKE images, a sufficiently long time interval (10 min) was inserted to make the domain structure in thermal equilibrium state after the temperature became stable. Figure 3(a) is the result for $V_G$ = 0 V, where the clear maze domain is observed. Then by applying positive and negative $V_G$ (= +10 and −10 V), the width of the domain was significantly expanded and narrowed[26,28] as shown in Figs. 3(b) and 3(c), respectively.

To determine the $V_G$ dependence of the domain width, the 2-dimensional fast Fourier transform (FFT) method was applied to the MOKE images[29] for an area of 180×180 μm². Figure 3(d) shows the averaged intensity of the FFTs for each direction of wavelength $d^{-1}$. From peak $d_p^{-1}$ in the spectrum, domain period $d_p$, which corresponds to averaged domain width $w_d^{ave}$, could be determined. $w_d^{ave}$ as a function of $V_G$ is shown in the inset of Fig. 3(d). For $V_G$ = +10(−10) V, $w_d^{ave}$ was determined to be 10(1.5) μm, whereas it was 6.5 μm at $V_G$ = 0 V. The ratio of the change in $w_d^{ave}$ from the value at $V_G$ = 0 V was +54 and −77% for $V_G$ = +10 and −10 V, respectively.

Domain width $w_d$ for an ultrathin ferromagnetic film is expressed as follows[27,30]:

$$w_d = 2\sqrt{\frac{A}{K_\perp^{eff}}} \exp\left[\frac{4\pi\sqrt{AK_\perp^{eff}}}{\mu_0 M_s^2 t_{Co}}\right] \quad (1),$$

where $K_\perp^{eff}$ is the effective perpendicular magnetic anisotropy energy constant, $M_s$ is the saturation magnetization, and $A$ is the exchange stiffness constant. So far, a large variation of domain width near $T_C$ has been ascribed to the reduction of $K_\perp^{eff}$, because the reduction of $K_\perp^{eff}$ causes the spin reorientation transition, resulting in an exponential decrease of the domain width.[31] Considering that the electric field also modulates the $K_\perp^{\text{eff}}$ near $T_C$,[19] the variation of $K_\perp^{\text{eff}}$ by EF may be the origin of the observed giant domain width modulation. We first checked this point. The variation of $K_\perp^{\text{eff}}$ (= $\mu_0 M_s H_s/2$) by EF was determined from saturation magnetic field $\mu_0 H_s$ of the hard-axis magnetization curve and $M_s$. Figure 4 shows the $R_{\text{Hall}}$ v.s. $\mu_0 H_{//}$ (in-plane magnetic field) curves for $V_G$ = +10, 0, and −10 V at 320 K. Using the relationship between normalized $R_{\text{Hall}}$ ($R_{\text{Hall}}^n$) and normalized in-plane magnetization $m_{//}$ (= $\sin[\cos^{-1}(R_{\text{Hall}}^n)]$), the hard-axis magnetization curve was obtained[19] as shown in the inset of Fig. 4, where $\mu_0 H_s$ (= $2\mu_0 \int_0^1 H_\parallel dm_\parallel$) was determined from the shaded area. In the inset, we excluded the data for $V_G$ = −10 V because of the formation of the multidomain state at $\mu_0 H_{//}$ = 0 T, and therefore, the $m_{//}$–$\mu_0 H_{//}$ curve could not be reproduced. This may indicate that a larger EF effect was induced at the negative $V_G$ side. From Fig. 4, the $\mu_0 H_s$ was measured to be $\mu_0 H_s$ = 1.14(1.09) T for 0 (+10) V. The variation of $M_s$ by EF was determined from the $M_s$ of the as-deposited sample (Fig. 2(a)) and the variation of $R_{\text{Hall}}$, thus $M_s$, by EF (Fig. 2(b)), and revealed to be 1.04(1.17) MA/m for 0 (+10) V. Based on the variation of $\mu_0 H_s$, and $M_s$, $K_\perp^{\text{eff}}$ was determined to be 0.59(0.64) MJ/m$^3$ for $V_G$ = 0 (+10) V. The variation of $K_\perp^{\text{eff}}$ by EF is about 8.5% and, based on Eq. (1), the resulting domain width variation is expected to be only about 2.7%. This is much smaller than the observed domain width variation by EF (54%). Note that the variation of $M_s$ also cannot account for the observed domain width variation because the positive voltage increases $M_s$ and $w_d$ in our experiment, which is contradict the prediction of Eq. (1).

The inconsistency naturally leads us to suspect the remaining term in Eq. (1), that is, the exchange stiffness $A$. Using the experimentally determined values of $K_\perp^{eff}$, $M_s$ and $t_{Co}$, $A$ is estimated based on Eq. (1) and is found to be $A$ = 0.12(0.18) pJ/m for $V_G$ = 0 (+10) V. Surprisingly, $A$ is increased by more than 50% for $V_G$ = +10V. Note that the result observed here ($V_G$ for the accumulation of electrons at the surface of the Co layer resulted in the increase of $A$ as well as $T_C$[10,11]) is consistent with the recent theoretical prediction.[25] The huge modulation of $A$ by EF is probably because the measurement temperature (320 K) was slightly below $T_C$ (= 347 K at $V_G$ = 0) where the EF effect became significantly enhanced. Lastly, we discuss reasons behind the quantitative disagreement of $A$ from the previous results. Although the significant variation of $A$ by EF is clear, the value of the obtained $A$ is more than one or two order smaller than the previous reported values.[32] We consider that the $A$ can be reduced drastically as the temperature approaches to $T_C$.[33] The drastic change of the bubble domain structure obtained slightly below 320 K (not shown), where the $K_\perp^{eff}$ and $M_s$ does not significantly vary with temperature, suggests the rapid decrease in $A$ with temperature near $T_C$. The small value of $A$ is also attributed to the ultrathin magnetic layer thickness of our film (1 monolayer of Co) because $A$ is generally decreased in ultrathin layer.[32] An in-plane non-uniformity of the magnetic parameters in such ultra-thin film, that results in the smaller domain width,[34] may make the effective *A* that obtained by Eq. (1) smaller. We also cannot rule out the possibility that the Eq. (1) cannot be applicable to our case since that equation is only valid for temperature much lower than $T_C$. Thus, the new model for ultra-thin ferromagnetic films that including the thermal energy may be required to quantitatively account the variation of domain width and exchange stiffness by EF around the $T_C$.

In summary, we observed a significant change in the domain size by applying an EF to a Pt/Co system. Although the EF changed the saturation magnetization and the magnetic anisotropy energy of the system, the change in the exchange stiffness was mainly attributed to the EF-dependent modulation of the domain width if the sample temperature was slightly below $T_C$.

The authors thank T. Dohi and F. Matsukura for useful discussion. This work was partly supported by JSPS KAKENHI (Grant Numbers 26103002, 25220604, 15H05702, 15H05419, and 2604316), the ImPACT Program of CSTI, and Collaborative Research Program of the Institute for Chemical Research, Kyoto University.

**Figure Captions**

**FIG. 1** Schematic illustration of the device structure and the measurement setup.

**FIG. 2** (a) Temperature dependence of the perpendicular component of magnetization $M_\perp$ of the as-deposited sample (left vertical axis). The right axis shows the temperature dependence of magnetic susceptibility $1/\chi_{\text{Hall}}$ determined from the Hall measurement for gate voltage $V_\text{G} = 0$. The Hall resistance $R_{\text{Hall}}$ curves for each temperature obtained by sweeping perpendicular magnetic field $\mu_0 H_\perp$ are shown in the inset. (b) The $R_{\text{Hall}}$ curves for $V_\text{G} = +10$, 0, and −10 V at 320 K. (c) The $R_{\text{Hall}}$ curves for $V_\text{G} = +10$, 0, and −5 V at 340, 335, and 330 K, respectively. The inset shows the magnified view of the data.

**FIG. 3** (a)–(c) Images taken using the magneto-optical Kerr effect (MOKE) microscope at 320 K if $V_\text{G} = 0$, +10, and −10 V was applied. (d) The averaged intensity of the 2-dimensional fast Fourier transforms (FFTs) applied to the MOKE images for each direction of wavelength $d^{-1}$.

**FIG. 4** The $R_{\text{Hall}}$ curves at 320 K obtained if in-plane magnetic field $\mu_0 H_{//}$ for $V_\text{G} = +10$, 0, and −10 V was applied. The inset shows the normalized in-plane magnetization curves for $V_\text{G} = +10$ and 0 V reproduced from the $R_{\text{Hall}}$ curves in the main graph.

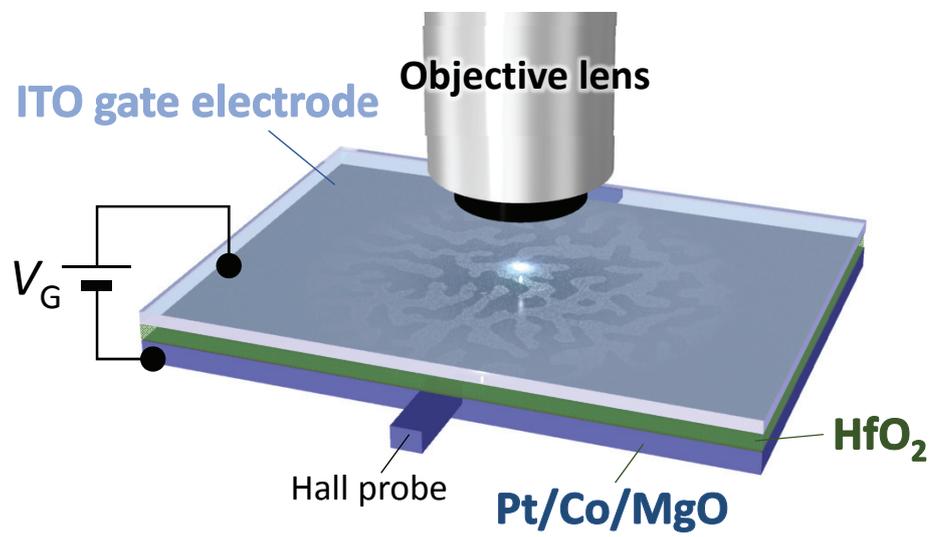

Ando *et al.*, FIG. 1

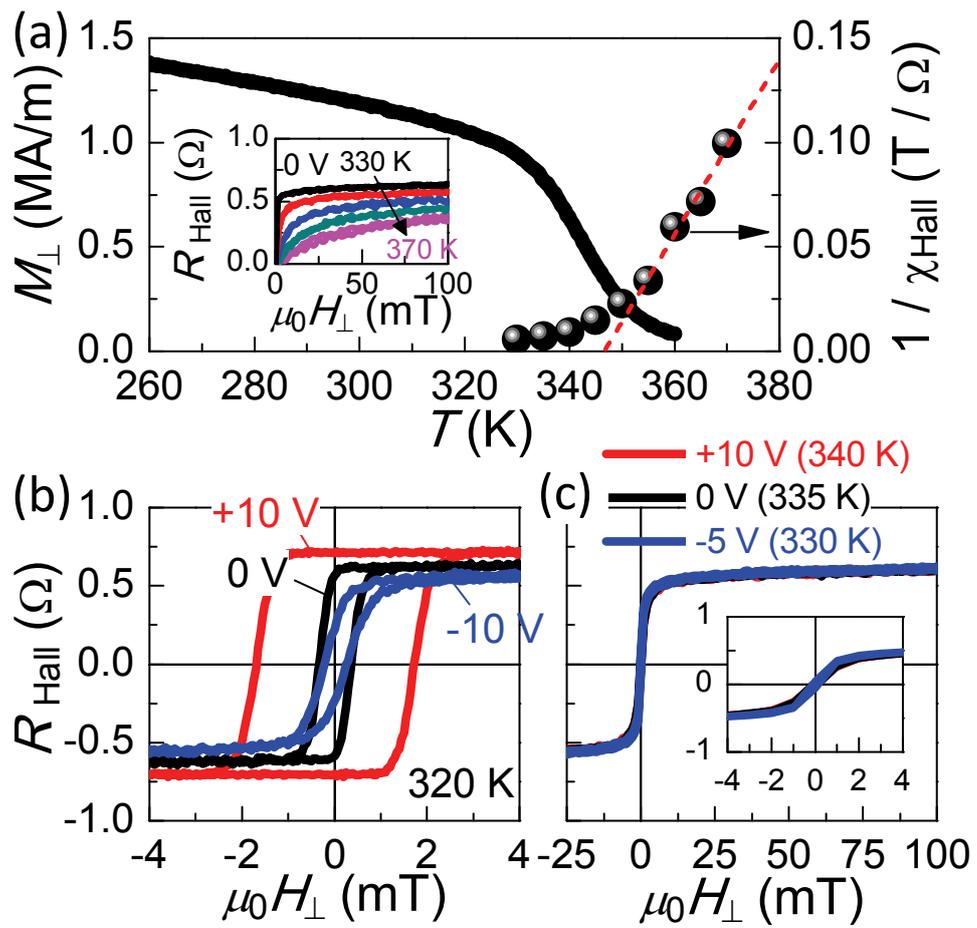

Ando et al., FIG. 2

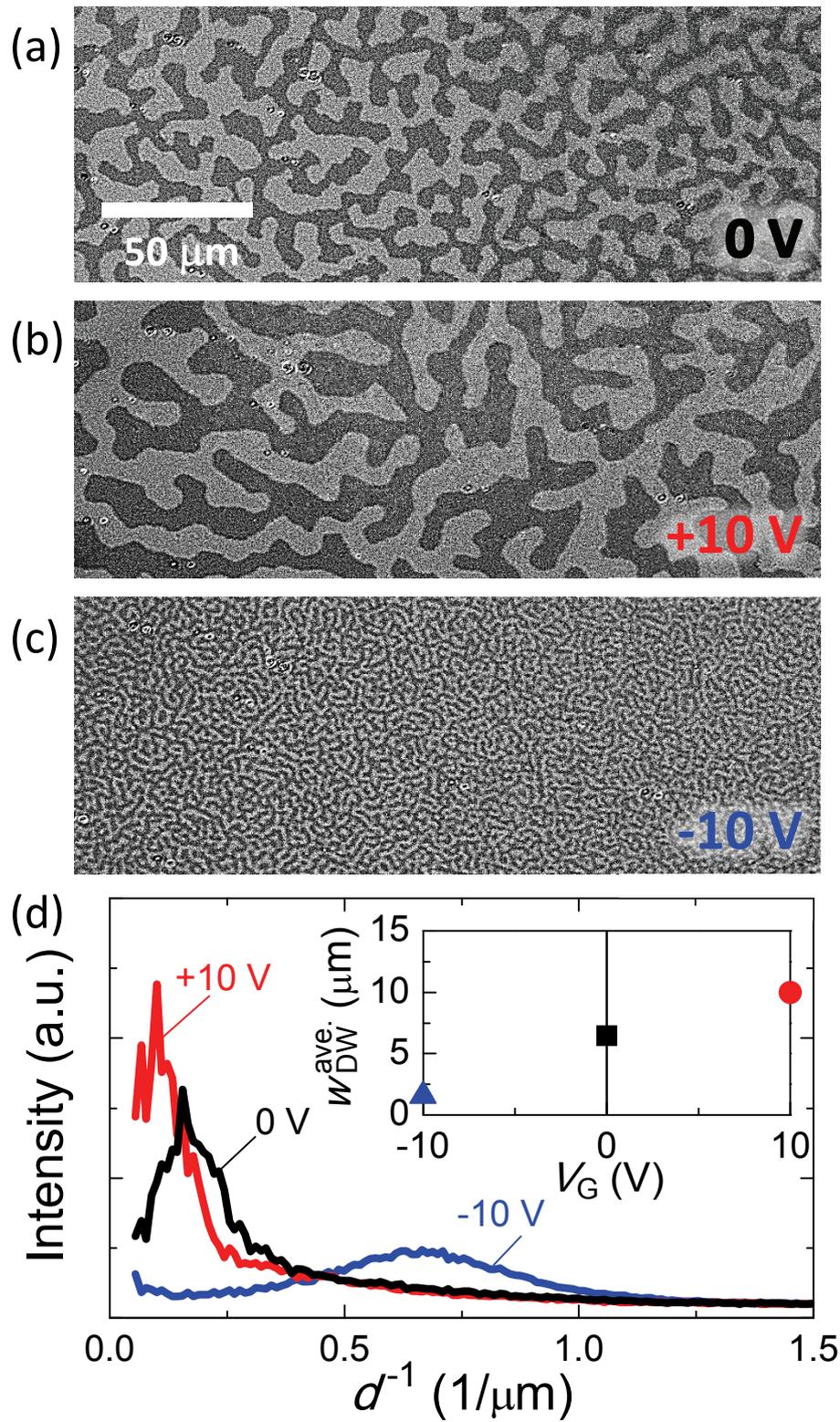

Ando et al., FIG. 3

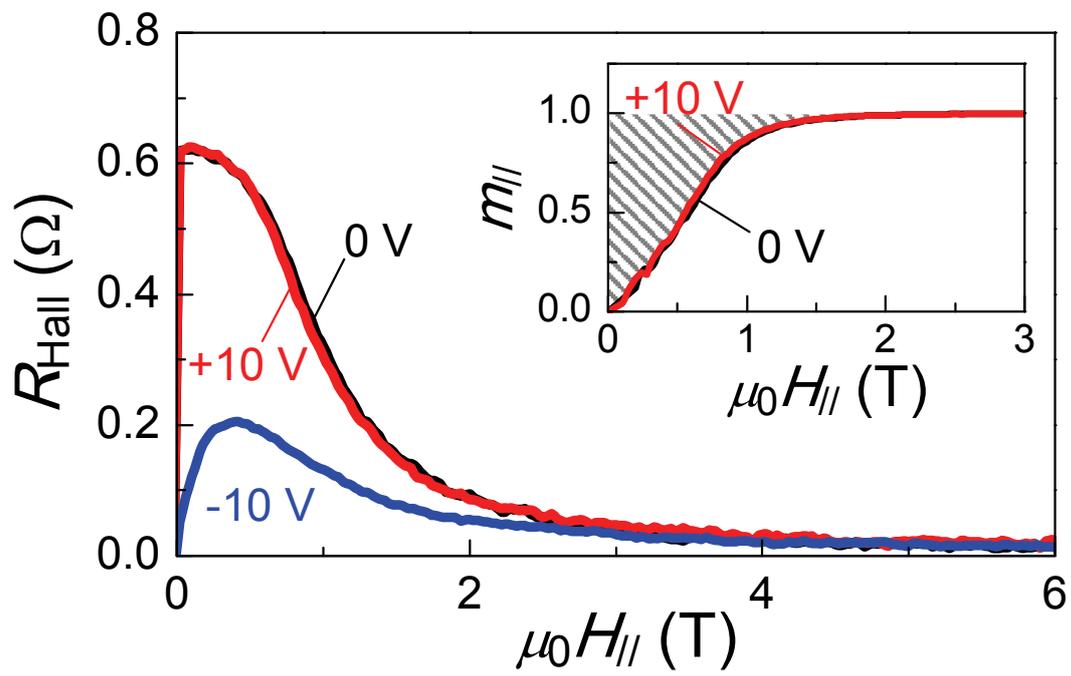

Ando et al., FIG. 4